\documentclass[preprint,showpacs,amsmath,amssymb]{revtex4}
\usepackage{graphicx}
\begin{document}

%
%
\title{Control of entanglement with multipulse application}  
\author{Chikako Uchiyama\(^{1}\) and Masaki Aihara\(^{2}\)}
\affiliation{%
\(^{1}\) Interdisciplinary Graduate School of Medicine and Engineering, 
University of Yamanashi,\\
4-3-11, Takeda, Kofu, Yamanashi 400-8511, JAPAN,\\
\(^{2}\)Graduate School of Materials Science,
Nara Institute of Science and Technology,\\
8916-5, Takayama-cho, Ikoma, Nara 630-0101 JAPAN
}%
\date{\today} 
%
\def\ch{{\cal H}}
\def\hbar{\mathchar'26\mkern -9muh}
\def\muv{{\vec \mu}}
\def\ev{{\vec E}}
\def\es{{\vec S}}

\def\ih{\frac{i}{\hbar}}
\def\taus{{\tilde \tau}_{s}}
\def\tauc{{\tilde \tau}_{c}}
\def\ee{|e\rangle \langle e|}
\def\eg{|e\rangle \langle g|}
\def\ge{|g\rangle \langle e|}
\def\gg{|g\rangle \langle g|}

\begin{abstract}
We show that the multipulse application can suppress the decoherence of quantum entanglement with focusing on the concurrence, the degree of entanglement.  By evaluating the time evolution of concurrence with a linearly interacting spin-boson model under pulse application, we find that the effectiveness of the multipulse control depends on the non-Markovian nature of the reservoir.    
\end{abstract}


\maketitle

The application of quantum principle to the field of information processing has opened a new perspective for secure communication and high-speed computation.  However, when we execute the schemes of quantum information processing, we have to overcome several obstacles which stem from the frailty of quantum substance.  Above all, the vulnerability of quantum entanglement to a noisy reservoir constitutes a serious obstacle to the realization of quantum teleportation\cite{bennett93} and quantum computation\cite{chuang00}. 

Since the methods to combat against the decoherence of the entanglement have been firstly proposed for the schemes of quantum teleportation\cite{bennett96} and quantum cryptography\cite{deutsch96}, many studies have been made under the condition/assumption that the entangled qubits are separated in space, and that we can only execute Local Operation and Classical Communication(LOCC) after the qubit has been contaminated by a noisy reservoir.  These studies, called as entanglement concentration, purification, or distillation\cite{bennett962,bennett961,duan00,bose99}, have tried to extract a smaller number of purely entangled pairs from two or more contaminated entangled pairs by applying LOCC. While previous experiments have realized the scheme of distillation\cite{kwiat,yamamoto} or purification\cite{zeilinger1}, it is necessary to use high accuracy measurement of a qubit (photon) for classical communication and an almost infinite number of contaminated pairs of qubits to obtain a perfectly pure entangled pair; a process which seems practically unfeasible.

In the field of quantum computation, many methods have been proposed to protect a single qubit against the effect of a noisy reservoir, which is called as decoherence.  These are roughly categorized into three types:(1) methods to obtain one stable qubit with ancillary qubits\cite{shor,symmetrize}, (2) methods to suppress the decoherence with quantum feedback\cite{vitali1}, and (3)application of pulse train(bang-bang or dynamical decoupling method)\cite{lloyd1,lloyd2}. Above all, the bang-bang or dynamical decoupling method has attracted much attention for its feasibility to various kinds of systems\cite{agarwal1}.  However, the physical background of the bang-bang method was not fully discussed in these studies.  In our recent work\cite{uchiyama1,uchiyama2}, we discussed the effectiveness of the bang-bang method on the decoherence of a single qubit and showed that the key is in the partial reversibility of qubit, which arises from a non-Markovian nature of decoherence process.  

When a qubit interacts with a reservoir that has a finite correlation time, the phase relaxation is described by an integral-differential equation for the off-diagonal element of density operator  (or induced moment ) \(P(t)\) of qubit as \( \frac{d}{{dt}}P(t) =  - \int_0^\infty  {dt'\;M(t')} P(t - t')\) which is obtained by the projection operator method\cite{ide}.  The memory kernel \(M(t')\) is described by correlation functions of reservoir and often has significant values only when time \(t'\) is shorter than the correlation time of reservoir \(\tau_{c}\). This indicates that the time derivative of \(P(t)\) depends on its past history, which we call non-Markovian nature.  When time goes much longer than \(\tau_{c}\), the integral-differential equation is approximated as \( \frac{d}{dt}P(t) =  - P(t) / T_{2} \) to obtain the exponential dephasing \( P(t) = e^{ - t/T_2 } P(0)\) where \(T_2\) is defined as \(T_2^{-1}  \equiv \int_0^\infty  {dt'\;M(t')} \).   The exponential decay, which is obtained by the above-mentioned Markovian approximation, indicates that the time evolution is irreversible.  Since the correlation time of reservoir \(\tau_{c}\) is usually much shorter than \(T_2\), the exponential dephasing is observed in various materials.   However, it should be noted that in the time scale which is comparable or shorter than \(\tau_{c}\), \(P(t)\) shows the non-exponential decay, which cannot be described by the constant dephasing time \(T_2\). In other words, during such time scale, the time evolution of \(P(t)\) is partially reversible.   The possibility to retrieve the reversibility from an induced moment with partial reversibility was firstly investigated in the scheme of optical transient four-wave mixing (optical parametric effect) \cite{aihara1,aihara2,fainberg,mukamel}  where application of  a \(\pi\)-pulse on an induced moment causes time-reversal action, and therefore \(P(t)\) is recovered depending on the degree of time-reversibility of the system.   This was verified by the experiments for sodium resorufin in dimethylsulfoxide\cite{nibbering}, iron-free myoglobin\cite{saikan}, dye molecules\cite{bigot}, and CdSe quantum dots\cite{woggon}. 

The bang-bang method using the \(\pi\)-pulse train can be understood as a natural extension of the two pulse optical transients mentioned above, as was pointed out in \cite{uchiyama1}.  That is, the repeatedly applied  \(\pi\) pulses with pulse interval much shorter than \(\tau_{c}\) can prevent the decoherence of qubit because of the non-Markovian nature of the system.   In the limit of zero pulse intervals in bang-bang method, the control of phase relaxation is possible even for the exponential dephasing.  However, it should be noted that we should take the zero limit of pulse interval \(\tau_{s}\) before we take the zero limit of \(\tau_{c}\); equivalently, the condition of \(\tau_{s}/\tau_{c} \rightarrow 0\) should be satisfied when we take the limit \(\tau_{s}\rightarrow 0\) \cite{lloyd2}.  Therefore, \(\tau_{c}\) should be comparable or larger than the pulse interval in any case, and the non-Markovian property of decoherence of qubit is essential for the effectiveness of bang-bang control.  

In this paper, we show that the multipulse application can control entanglement for a pair of qubits.   In order to analyze the controllability of entanglement by multipulse application, it is plausible to evaluate a quantity called concurrence \cite{wooters97} that plays a center role to describe the degree of entanglement of a pair of qubits.  For a density matrix \(\rho\) of a pair of qubits under consideration, the concurrence \(C(\rho)\) is defined by

\begin{equation}
C(\rho) = {\rm max} (0, 2{\lambda_{{\rm max}}} - {\rm Tr} R).
\label{eqn:1}
\end{equation} 
Here \(\lambda_{{\rm max}}\) is the maximum eigenvalue of the operator \(R\)  which is defined by
\begin{equation}
R = \sqrt {\sqrt{\rho} {\tilde \rho} \sqrt{\rho}},
\label{eqn:2}
\end{equation} 
for the density matrix of the pair of qubits \(\rho\) and \({\tilde \rho}\),
\begin{equation}
{\tilde \rho} = (\sigma_{1,y} \otimes \sigma_{2,y}) \rho^{*} (\sigma_{1,y} \otimes \sigma_{2,y}),  
\label{eqn:3}
\end{equation} 
where the \(\sigma_{n,y}\) indicates \(y\)-component of the Pauli matrix of \(n\)-th qubit and \(\rho^{*}\) means the complex conjugate of density matrix \(\rho\).  Since the product of \(\sigma_{n,y}\) and complex-conjugated operator is associated with time-reversal operator\cite{sakurai}, the concurrence is based on the ``degree of equality"\cite{wooters97,bures} between a density matrix under consideration and a density matrix obtained by time-reversal operation.  By this definition, we can consider that the concurrence describes the degree of reversibility.  This is exactly the reason why the concurrence is an appropriate quantity when we consider the \(\pi\) pulse application to a pair of qubits that interact with a reservoir of non-Markovian nature, since the pulse application causes time-reversal in the evolution of the qubits.
 
Taking a simple case where a pair of qubits linearly interact with a common reservoir, let us show the effectiveness of multipulse application to suppress the disentanglement. The Hamiltonian of this system is 
\begin{equation}
\ch_{R}= \ch_{0} + \ch_{SB} = ( \ch_{S} + \ch_{B} )+ \ch_{SB}\;,
\label{eqn:4}
\end{equation}
with \( \ch_{S}= \hbar \sum_{n=1}^{2} \omega_{0} S_{n,z}  \), \(\ch_{B}=\sum_{k} \hbar \omega_{k} b_{k}^{\dagger} b_{k} \), and 
\begin{equation}
\ch_{SB} = \hbar \sum_{n=1}^{2}  S_{n,z}  \sum_{k} h_{k} \omega_{k}(b_{k}+ b_{k}^{\dagger}). 
\label{eqn:5}
\end{equation} 
Here  \(S_{n,z} \) is the \(z\)-component of the \(n\)-th qubit \((n=1,2)\); \( b_{k} (b_{k}^{\dagger}) \) indicates the annihilation (creation) operator of the \(k\)-th boson which composes the reservoir; \(h_{k}\) is the coupling strength between the qubit; the \(k\)-th boson of the reservoir.  It is noted that the reservoir is common for two qubits, and the model is applicable to a pair of quantum dots in semiconductors\cite{braun}. To suppress the disentanglement due to the interaction with the reservoir, we apply sufficiently short \(\pi\)-pulse train on both qubits simultaneously. The Hamiltonian under pulse application is written as
\begin{eqnarray}
\ch_{SP}(t) &=& \ch_{S} + \sum_{j=0}^{N} \ch_{P,j}(t), \label{eqn:6} \\
\ch_{P,j}(t) &=&-\frac{1}{2} \ev_{j}(t) \cdot \muv \; \sum_{n=1}^{2} (S_{n,+} e^{-i \omega_{0} t}+S_{n,-} e^{i \omega_{0} t}) \label{eqn:7} 
\end{eqnarray} 
where \(\ev_{j}(t)\) denotes the field amplitude of \(j\)-th applied pulse.  We assume that two qubits have the same transition moment \(\muv \). When we consider a \(\frac{1}{2}\) spin (a two-level system) as a qubit, Eq.(\ref{eqn:6}) means the magnetic interaction (the electric interaction for optical transition), respectively .  

When we apply  \(N\) \(\pi\)-pulses with pulse interval \(\tau_{s}\) and pulse duration \(\Delta t\), the density operator for the total system \(\rho(t)\) is written as
\begin{eqnarray}
\rho(t)   &=& e^{-i L_{R} \times (t-N\tau_{s})} \nonumber \\
&&\hspace{-1cm}  \times \{\prod_{j=1}^{N} 
 e^{-i \int_{ j \tau_{s}-\Delta t}^{j \tau_{s} } dt' L_{SP} (t')}  e^{-i L_{R} \times (\tau_{s}-\Delta t)} \}
\rho(0) , 
\label{eqn:8}
\end{eqnarray}
where \(\rho(0)\) indicates the density operator for the total system at an initial time \(t=0\), and \(L_{R}\) and \(L_{SP}(t)\) are Liouville operators defined by \(L_{R} X \equiv \frac{1}{\hbar} [\ch_{R},X] \) and \(L_{SP}(t) X \equiv \frac{1}{\hbar} [\ch_{SP}(t),X] \) for an arbitrary operator \(X\). The exponential of Liouville operators are related to the exponentials of the corresponding Hamiltonians as \(e^{-i L_{R} t} X = e^{-\frac{i}{\hbar} \ch_{R} t} X e^{\frac{i}{\hbar} \ch_{R} t} \) and \(e^{-i \int_{ j \tau_{s}-\Delta t}^{j \tau_{s} } dt' L_{SP} (t')} X = e^{-\frac{i}{\hbar} \int_{ j \tau_{s}-\Delta t}^{j \tau_{s} } dt' (\ch_{S}+\ch_{P,j} (t'))} X e^{ \frac{i}{\hbar} \int_{ j \tau_{s}-\Delta t}^{j \tau_{s} } dt' (\ch_{S}+\ch_{P,j} (t'))} \) where we assumed that the pulses are well separated.

In order to evaluate the time evolution of the density operator, it is convenient to use a relation for the time evolution operator as
\(e^{-\frac{i}{\hbar} \ch_{R} t} = e^{-\frac{i}{\hbar} \ch_{0}  t} T_{+} \exp[-\frac{i}{\hbar} \int_{0}^{t} {\tilde \ch}_{SB}(t') dt'] \)
where \(T_{+}\) denotes the time ordering symbol from right to left, and \({\tilde \ch}_{SB}(t)  \equiv e^{\frac{i}{\hbar} \ch_{0} t} \ch_{SB} e^{-\frac{i}{\hbar} \ch_{0} t}\).  When we consider the decoherence caused by Eq.(\ref{eqn:5}), the time ordered exponential is written as 
\begin{equation}
T_{+} \exp[-\frac{i}{\hbar} \int_{0}^{t} {\tilde \ch}_{SB}(t') dt']  = {\rm Diag} [u_{+}(t),1,1,u_{-}(t)],
\label{eqn:9}
\end{equation}
where \({\rm Diag}[ \dots ]\) means a diagonal matrix with elements in the bracket \([ \cdots ]\), and \(u_{\pm}(t) \equiv T_{+} \exp[ \mp \frac{i}{\hbar} \int_{0}^{t} B(t') dt'] \) with \(B(t) \equiv \hbar \sum_{k} h_{k} \omega_{k}(b_{k} e^{-i \omega_{k} t} + b_{k}^{\dagger} e^{i \omega_{k} t}) \).  Assuming that we apply square pulses with height \(\ev_{j}\) for the \(j\)-th pulse, we obtain the time evolution operator under pulse application as 
\(e^{-\frac{i}{\hbar}  \int_{ j \tau_{s}-\Delta t}^{j \tau_{s} } dt' (\ch_{S}+\ch_{P,j} (t'))}= e^{-\frac{i}{\hbar} \ch_{S}  \Delta t } e^{-\frac{i}{\hbar} {\tilde \ch}_{P,j}  \Delta t } \), where  \({\tilde \ch}_{P,j}\) means the pulse Hamiltonian in the interaction picture with \({\tilde \ch}_{P,j} \equiv e^{\frac{i}{\hbar} \ch_{S} t} \ch_{P,j} (t) e^{-\frac{i}{\hbar} \ch_{S} t} = -\frac{1}{2} \ev_{j} \cdot \muv \; \sum_{n=1}^{2} (S_{n,+} +S_{n,-})\).  

Let us consider the following initial conditions for the qubits and the boson reservoir: (1) the two qubits maximally entangled at an initial time as \(|\psi\rangle=\frac{1}{\sqrt{2}}(|1\rangle|1\rangle +|0\rangle|0\rangle) =\frac{1}{\sqrt{2}}(|11\rangle +|00\rangle)\) where \(|0\rangle\) and \(|1\rangle\) indicate the two states of a qubit, and (2) the boson reservoir is in the vacuum state.  Then we have the density operator at the initial time as
\begin{equation}
\rho(0)=|\psi\rangle \langle \psi | \otimes \rho _{R} = \frac{1}{2}\left( {\begin{array}{*{20}c}
   {\rho _R } & 0 & 0 & {\rho _R }  \\
   0 & 0 & 0 & 0  \\
   0 & 0 & 0 & 0  \\
   {\rho _R } & 0 & 0 & {\rho _R }  \\
\end{array}} \right),
\label{eqn:10}
\end{equation}
where  \(\rho _{R} =| 0 \rangle \langle 0 | \) is the density operator for the boson reservoir.

Using Eq.(\ref{eqn:1}) \(\sim\) Eq.(\ref{eqn:10}), we obtain the time evolution of concurrence under \(N\)  \(\pi\)-pulses with pulse interval \(\tau_{s}\) in the limit of infinitely short pulse duration \(\Delta t \rightarrow 0 \) as follows:
\begin{equation}
C(t)= \frac{1}{2} \exp[-2 \sum_{k} |\alpha _{k} (t)|^2],
\label{eqn:11}
\end{equation} 
where 
\begin{eqnarray}
\alpha _{k} (t) &=& h_k e^{ - i\omega _k (t - N\tau _s )} \{( {1 - e^{i\omega _k (t - N\tau _s )} }) \nonumber \\
&& + \sum_{m = 1}^{N} {(- 1)^m e^{ - i m \omega _k \tau _s } } (1 - e^{  i\omega _k \tau _s } ) \}.
\label{eqn:12}
\end{eqnarray} 

Equation (\ref{eqn:11}) is similar to the formula that we have obtained for the decoherence of a single qubit under linear interaction with the boson reservoir(see Eq.(29) in \cite{uchiyama2}). The exponent in Eq.(\ref{eqn:7}) is twice larger than that in Eq.(29) in \cite{uchiyama2} which is the signal intensity expressed by the off-diagonal element of the reduced density matrix as \( I(t)=| \langle e| {\rm Tr}_{R} \rho(t) | e \rangle |^2 \) where \(| e \rangle (| g \rangle)\) corresponds to \(| 0 \rangle (| 1 \rangle)\), respectively.  This means that the entanglement of two qubits decays four times faster than the single qubit, since the concurrence is given by the off-diagonal element of density matrix of two qubits as \( |  \langle 0 0| {\rm Tr}_{R} \rho(t) | 1 1 \rangle |^2 \). The same result has been pointed out by Yu and Eberly\cite{eberly}.  

To evaluate Eq.(\ref{eqn:11}), we need an explicit form of the coupling function
\begin{equation}
h(\omega) \equiv \sum_{k} |h_{k}|^2 \delta(\omega-\omega_k).  
\label{eqn:12}
\end{equation} 
Bang-bang control is effective for any type of coupling function when the pulse train is applied with infinitely small pulse interval, which is practically difficult to execute. As we discussed for the decoherence of a single qubit\cite{uchiyama2}, we can release such a difficulty by paying attention to the dynamical behavior of reservoir: by synchronizing the pulse-train application with the characteristic oscillation period of the reservoir, we can efficiently suppress the decoherence of the qubit, a process which we call ``synchronized pulse control" (SPC)\cite{uchiyama2}.  The effectiveness of pulse control with finite pulse interval, especially for SPC, strongly depends on the type of coupling function.  Fixing the width of the coupling function at a relatively small value, we compared the effectiveness for Lorentzian and non-Lorentzian coupling function.  We found that SPC is more effective for non-Lorentzian coupling function than Lorentzian coupling function. We can explain the reason with a picture of two-step structured reservoir which is transformed from the original spin-boson model.  When the coupling function is Lorentzian,  we can reconstruct the reservoir which consists of infinite number of bosons to obtain a single boson, called an interaction mode\cite{toyozawa} or quasi mode\cite{lang}, which interacts with another reservoir of infinite number of bosons with a flat (white) coupling function.  In this case, the interaction mode shows the Markovian time evolution.   When a single qubit interacts with such a boson with Markovian nature, the qubit loses partial reversibility quickly, which makes the pulse control less effective.  Since Eq.(\ref{eqn:11}) is similar to the time evolution of the decoherence of a single qubit under pulse application, we can say that the pulse control on entanglement is also less effective for Lorentzian coupling function.   To discuss the effectiveness of pulse control with finite pulse interval, let us consider the non-Lorentzian coupling function. 

As a typical example of non-Lorentzian coupling function, we assume the coupling function to be a Gaussian distribution with the mean frequency \(\omega_{p}\) and the variance \(\gamma_{p}\),
\begin{equation}
h(\omega) \equiv \frac{s }{\sqrt{\pi} \gamma_{p} } \exp(-\frac{(\omega-\omega_{p})^2}{\gamma_{p}^2}),
\label{eqn:13}
\end{equation}
where \(s\) is the average number of bosons interacting with a qubit, which means the strength of coupling between a qubit and reservoir.   

In Fig.1, we show the time dependence of the concurrence \(C({\tilde t})\), wherein a scaled time variable \( {\tilde t} \equiv t \omega_{p}\) is used, and the parameters are set as \({\tilde \gamma_p\equiv \gamma_p/\omega_p}=0.1 \) and \( s= 5\). Without pulse application, one can see that the concurrence shows a damped oscillation with the center frequency \(\omega_{p}\) of the coupling function (the mean period is \(2\pi\) in the scaled time) as shown in Fig.1(a). In the figure, we can see that the partial reversibility of concurrence remains during the mean period of dynamical motion of boson reservoir.  When one applies a \(\pi\)-pulse train with a relatively short interval \(\tilde{\tau}_{s}=\pi/5 \), one can see that the damping of the concurrence is reduced and approaches to a constant value(Fig.1(b)). This result clearly indicates that the degradation of concurrence is effectively suppressed by the application of \(\pi\)-pulse train.   With increasing the pulse interval, the concurrence decreases.  At the pulse interval \(\tilde{\tau}_{s}=\pi\), the concurrence takes even smaller values than those in the case of no pulse control as shown in Fig.1(c).  However, when the pulse application is synchronized with the oscillation period of the center frequency of the coupling function by setting \(\tilde{\tau}_{s}=2 \pi\), one obtains a periodically recovering concurrence.  The recovery means that the time-reversal action by \(\pi\)-pulse application efficiently recover the coherence of the quantum entanglement by paying attention to the dynamical motion of the boson reservoir which causes the memory (non-Markovian) effect on the time evolution of concurrence.      

\begin{figure}[h]
\begin{center}
\includegraphics[scale=0.5]{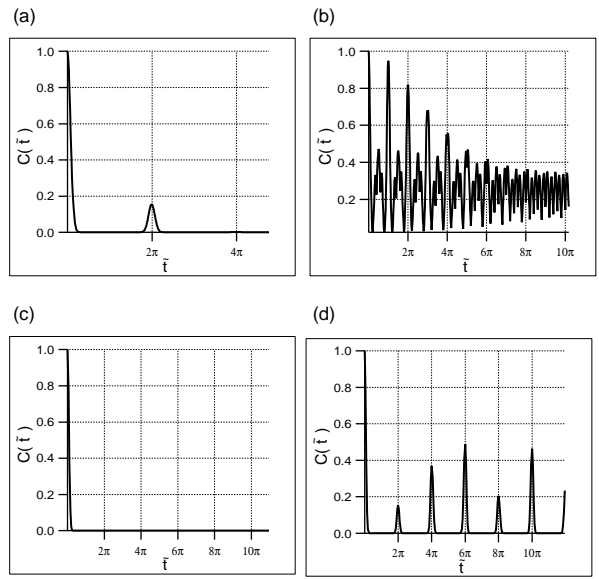}
\end{center}
\caption{Time evolution of \(C({\tilde t} )\) for \({\tilde \gamma_p \equiv \gamma_p/\omega_p} =0.1\), \( s= 5\). (a) shows the case without pulse application, (b) for pulse interval 
\({\tilde \tau}_{s}=\pi/5\), (c) for \({\tilde \tau}_{s}=\pi\), (d) for \({\tilde \tau}_{s}=2 \pi\).} 
\label{fig:fg1}
\end{figure}

As a measure of entanglement, we use the concurrence in this paper. When the interaction between the qubits and the reservoir is written as Eq.(\ref{eqn:2}), we can relate other measures with concurrence. Firstly, we consider the entropy of the pair of qubits \(S(t) = - \sum_{i=1}^{4} \lambda_{i} \log_{4} \lambda_{i}\) which has been proposed by Thew and Munro \cite{thew}. They categorized the function of entanglement distillation, purification, and concentration by the dependence of the entanglement of formation(EOF) on the entropy \(S(t)\). They defined the concentration as the procedure to decrease the entropy and increase EOF, purification as that to decrease entropy without changing EOF, and distillation as that to increase EOF without changing entropy.  For the present model where the system-reservoir interaction is adiabatic, we obtain a simple relation between the time evolution of entropy and the concurrence as \(S(t) = - \sum_{i=1}^{4} \lambda_{i} \log_{4} \lambda_{i}\) where \(\lambda_{1}=(\frac{1+C(t)}{2})^2, \lambda_{2}=(\frac{1-C(t)}{2})^2, \lambda_{3}=\lambda_{4}=0\).    Since this relation holds under the pulse excitation, the concurrence and the entropy  \(S(t)\) are mutually related during the course of entanglement recovery by multipulse application, as Fig.2 shows. The entropy \(S(t)\) ranges between \(0\) and \(0.5\) in this case.  This is because the density matrix of the most phase decohered state has only two elements as \( \langle 00 | Tr_{R}\rho(t) | 00 \rangle =\langle 11 | Tr_{R}\rho(t) | 11 \rangle =\frac{1}{2}  \) .  We also found the relation between the purity of qubits \(P(t)=Tr(\rho(t)^2)\) and the concurrence as \(P(t)=\frac{1+C(t)^2}{2}\), which shows how the multipulse application can recover the purity of the qubits as the concurrence decreases.
\begin{figure}[h]
\begin{center}
\includegraphics[scale=0.5]{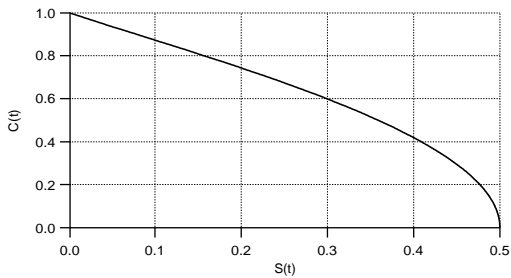}
\end{center}
\caption{The dependence of \(C(t)\) on \(S(t)\).}
\label{fig:fg2}
\end{figure}

We also consider the case in which each qubit individually interacts with its own boson reservoir:
\begin{equation}
\ch_{SB} = \hbar \sum_{n=1}^{2}  S_{n,z}  \sum_{k_{n}} h_{k_{n}} \omega_{k_{n}} (b_{k_{n}}^{(n)} + {b_{k_{n}}^{(n)}}^{\dagger}). 
\label{eqn:14}
\end{equation} 
Here \(b_{k_{n}}^{(n)}\)  \(({b_{k_{n}}^{(n)}}^{\dagger})\) indicates the creation (annihilation) operator of boson reservoir with which \(n\)-th qubit interacts.  In Eq.(\ref{eqn:14}),  \(\omega_{k_{n}}\) is the frequency of the \(k\)-th boson of the \(n\)-th reservoir. \(h_{k_{n}}\) indicates the coupling strength between the \(n\)-th qubit and the \(k\)-th boson.  
At an initial time, we assume the qubits to be in the maximally entangled state as before. In this case, we obtain the expression for the time evolution of concurrence as,
\begin{equation}
C(t)= \frac{1}{2} \prod_{n=1}^{2} \exp[- \frac{1}{2} \sum_{k_n} |\alpha _{k_n} (t)|^2],
\label{eqn:15}
\end{equation} 
where \(\alpha _{k_n} (t)\) is given by replacing suffixes \(k\) in Eq.(\ref{eqn:12}) with \(k_n\).  When we set the same coupling function for two reservoirs, \({\tt ln} C(t)\) in non-common reservoir case is a half of that in the common reservoir case in Eq.(\ref{eqn:11}).  This means that the qualitatively the same effect of a \(\pi\)-pulse train is observed for non-common reservoir case.  
 Consequently, \({\tt ln} C(t)\) becomes a half of that in the case of common reservoir shown in Eq.(\ref{eqn:7}), if we set the same coupling function for two reservoirs.  This is the only difference between the two cases of common and non-common reservoirs. Qualitatively the same effect of a \(\pi\)-pulse train is observed.  

In this paper, we have analyzed the case where the two qubits do not interact each other directly, although the interaction between qubits is necessary to generate entanglement.   This is because, when we assume the initial state of the pair of qubits to be  \(|\psi\rangle=\frac{1}{\sqrt{2}}(|11\rangle +|00\rangle)\), the Heisenberg interaction as \( \es \cdot \es \) does not affect the time evolution of concurrence.  Then, the interaction with common or non-common reservoir is sufficient to discuss the disentanglement in this case.   

In the evaluation of concurrence, we have used the Gaussian coupling function as an example.  Since the formula of concurrence is found to be similar to the formula for a single qubit\cite{uchiyama2}, we can say that the effectiveness of pulse control on the concurrence is also more effective for semi-elliptic coupling function than the Gaussian one, especially for the synchronized pulse control(SPC).  This is because the semi-elliptic function does not have a tail, and is very different from the Lorentzian shape.  However, when the pulse interval is infinitely small, the bang-bang control is effective for any type of coupling function.  

In summary, we have shown that the multipulse application can suppress the decoherence of quantum entanglement with focusing on the concurrence in the non-common reservoir case as well as in the common reservoir case.  We found that the effectiveness of multipulse control depends on the non-Markovian nature of the reservoir.    

This study was supported by the Sumitomo Foundation and the Grant in Aid for Scientific Research from the Ministry of Education, Science, Sports and Culture of Japan. The authors are thankful to Prof. E.C.G.Sudarshan, Anil Shaji, and Cesar Rodriguez for their helpful comments and suggestions. The authors are also grateful to Lorenza Viola, Daniel Lidar, Mikio Nakahara, Shogo Tanimura and Fumiaki Morikoshi for their thoughtful discussions.

\end{document}